\documentclass{jpconf}
\usepackage{graphicx}

\makeatletter
\def\alt{\mathrel{\mathpalette\vereq<}}
\def\vereq#1#2{\lower3pt\vbox{\baselineskip1.5pt \lineskip1.5pt
\ialign{$\m@th#1\hfill##\hfil$\crcr#2\crcr\sim\crcr}}}
\def\agt{\mathrel{\mathpalette\vereq>}}
\makeatother

\long\def\dump#1{}

\begin{document}
\title{Supernova neutrino observations:\\ What can we learn?}

\vskip-12pt \vbox to 0pt{{\ }\newline\vskip-5cm \noindent
\normalsize\rm Contribution to Proc.\ {\it Neutrino 2006: XXII
International Conference on Neutrino Physics and Astrophysics},
13--19 June 2006, Santa Fe, New Mexico.\vfil}

\author{Georg G.~Raffelt}

\address{Max-Planck-Institut f\"ur Physik
(Werner-Heisenberg-Institut)\\
F\"ohringer Ring 6, 80805 M\"unchen, Germany}

\ead{raffelt@mppmu.mpg.de}

\begin{abstract}
Twenty years after SN~1987A, the vast international programme of
experimental neutrino physics and neutrino astronomy suggests that
large detectors will operate for a long time. It is realistic that a
high-statistics neutrino signal from a galactic SN will be observed.
I review some of the generic lessons from such an observation where
neutrinos largely play the role of astrophysical messengers. In
principle, the signal also holds valuable information about neutrino
mixing parameters. I explain some recent developments about the
crucial importance of collective neutrino oscillations in the SN
environment.
\end{abstract}

\section{Introduction}                        \label{sec:introduction}

Twenty years ago, the neutrino burst from supernova (SN) 1987A in
the Large Magellanic Cloud was observed. On 23~February 1987 at
7:35~h universal time, the Kamiokande~II~\cite{Hirata:1987hu,
Hirata:1988ad} and IMB~\cite{Bionta:1987qt, Bratton:1988ww}
water-Cherenkov detectors each registered a burst clearly attributed
to SN~1987A. A~contemporaneous signal in the Baksan scintillator
detector~\cite{Alekseev:1987ej, Alekseev:1988gp} may have been
caused by the neutrino burst as well. A significant event cluster in
the LSD experiment~\cite{Dadykin:1987ek, Aglietta:1987it} was
observed several hours earlier and had no counterpart in the other
detectors and vice versa. It can be associated with SN~1987A only if
one invokes very non-standard double-bang scenarios of stellar
collapse~\cite{Imshennik:2004iy}. A lively account of the exciting
and somewhat confusing history of the SN~1987A neutrino detection
was given by M.~Koshiba~\cite{Koshiba:1992yb} and
A.~Mann~\cite{Mann:1997}.

This unique observation of stellar-collapse neutrinos helped to pave
the way for a new era of neutrino physics. Today, the discovery of
neutrino masses, lepton mixing, and flavor oscillations are quickly
fading to become yesterday's sensation while the experimental
efforts are turning to yet more challenging issues, notably the
question of leptonic CP violation, the absolute neutrino masses, and
their Majorana nature. A broad programme of experimental neutrino
physics, dedicated SN neutrino observatories, and the construction
of IceCube as a high-energy neutrino observatory almost guarantee
the operation of large detectors for a long time so that the
eventual observation of a high-statistics SN neutrino burst is a
realistic possibility. A review of the ongoing, planned or proposed
neutrino experiments with SN detection capabilities was given by
K.~Scholberg at this conference~\cite{Scholberg:2007nu}.

In our galaxy, the SN rate is perhaps 1--3 per century, so that the
observation of a SN neutrino burst is a once-in-a-lifetime
opportunity. What can we learn? There is no simple answer to this
question because what we will learn depends on the detectors
operating at that time, what they will observe, what else we then
know about neutrinos, and which non-neutrino observations will be
available. Galactic SNe are typically obscured, but even then
probably would be seen, for example, in x- or $\gamma$-rays.
Moreover, a gravitational wave signal could be observed.

Forecasting all possible scenarios would be both impossible and
moot. Rather, I will focus on a number of generic issues. First, in
Sec.~\ref{sec:rate} I review current estimates of the galactic SN
rate and about their distance distribution. In
Sec.~\ref{sec:lessons} I will review some of the obvious lessons
from a SN neutrino observation. Here, neutrinos largely play the
role of astrophysical messengers. In
Sec.~\ref{sec:flavor-oscillations} I turn to flavor oscillations
where the observations could reveal crucial information about
neutrino mixing parameters. Until recently, the impact of collective
neutrino oscillations in the SN context had been underestimated.
Therefore, the overall picture of SN neutrino oscillations is in a
state of flux. Sec.~\ref{sec:summary} is given over to a summary and
conclusions.

\section{Next supernova: Where and When?}             \label{sec:rate}

Existing and near-future neutrino detectors~\cite{Scholberg:2007nu}
do not reach beyond the galaxy and its satellites. Super-Kamiokande
would observe about $10^4$ events from a SN at a typical galactic
distance of 10~kpc. The next significant target would be the
Andromeda region at a distance of 760~kpc, reducing the rate by
$(10/760)^2=1.7\times10^{-4}$, i.e., Super-K would register 1--2
events. If a megatonne detector is built with perhaps 30 times the
Super-K fiducial volume, it would provide several tens of events.
Even such a low-statistics observation would be very useful as we
shall see below. From the nearest galaxies beyond Andromeda, even a
megatonne detector would register only 1--2 events. It was noted,
however, that correlating them with astronomical SN observations may
allow one to reduce background enough to build up SN neutrinos at a
rate of perhaps 1~neutrino per year from galaxies out to
several~Mpc~\cite{Ando:2005ka}.

One classic method to estimate our galaxy's SN rate is to scale from
external galaxies. Another classic approach is to extrapolate the
five historical SNe of the past millenium to the entire galaxy,
leading to a larger but more uncertain estimate. The most recent
estimate derives from the $\gamma$-rays emitted by $^{26}$Al
(half-life $7.2\times10^5$ years) that is produced in massive stars.
Finally, the non-observation of a galactic neutrino burst since
30~June~1980 when the Baksan Scintillator Telescope (BST) took up
operation, and the almost complete coverage of the neutrino sky by
different detectors since then, provides the upper limit shown in
Table~\ref{tab:rate}.

\begin{table}[b]
 \caption{\label{tab:rate}Estimated rate of galactic core-collapse
 SNe per century.}
 \centering
 \begin{tabular}{llll}
 \br
 Method&Rate&Authors&Refs.\\
 \mr
 Scaling from external galaxies&$2.5\pm0.9$
 &van den Bergh \& McClure&
 \cite{vandenBergh:1994, Diehl:2006cf}\\
 &&(1994)\\
 &$1.8\pm1.2$&Cappellaro \&\ Turatto&
 \cite{Cappellaro:1999qy, Cappellaro:2000ez}\\
 &&(2000)\\
 Gamma-rays from galactic $^{26}$Al&$1.9\pm1.1$
 &Diehl et al.\ (2006)&
 \cite{Diehl:2006cf}\\
 Historical galactic SNe (all types)&$5.7\pm1.7$&
 Strom (1994)&\cite{Strom:1994}\\
 &$3.9\pm1.7$&
 Tammann et al.\ (1994)&\cite{Tammann:1994ev}\\
 No neutrino burst in 25 years$^{a}$&${}<9.2$ (90\% CL)&
 Alekseev \& Alekseeva&\cite{Alekseev:2002ji}\\
 &&(2002)\\
 \br
 \multicolumn{4}{l}{$^a$We have scaled the limit of
 Ref.~\cite{Alekseev:2002ji} to 25~years of neutrino sky coverage.}
 \end{tabular}
\end{table}

Therefore, one expects 1--3 core-collapse SNe per century in our
galaxy and its satellites. With a megatonne-class detector one would
reach Andromeda (M31) and its immediate neighbors such as Triangulum
(M33), roughly doubling the expected rate. On the other hand, the
last SN from that region was observed in 1885! However, we also note
that SNe can be quite frequent in some galaxies. The record holders
are NGC~6946 with SNe 1917A, 1939C, 1948B, 1968D, 1969P, 1980K,
2002hh and 2004et and the galaxy NGC~5236 (M83 or Southern Pinwheel)
with SNe 1923A, 1945B, 1950B, 1957D, 1968L and 1983N~\cite{asiago}.
These time sequences provide a healthy lesson in Poisson statistics:
even if the average rate is quite large, one may still wait for a
long time for the next SN, or conversely, we could be lucky and
observe one soon, even if the average rate is as small as suggested
by Table~\ref{tab:rate}.

What would be a typical distance for a SN in our own galaxy?
Core-collapse marks the final evolution of massive stars and thus
must occur in regions of active star formation, i.e., in the spiral
arms. As proxies for the distribution one can use either
observations in other galaxies or in our galaxy the distribution of
pulsars, SN~remnants, molecular and ionized hydrogen, and OB-star
forming regions~\cite{Ferriere:2001rg}. All of these observables are
consistent with a deficit of SNe in the inner galaxy and a maximum
at 3.0--5.5~kpc galactocentric distance. Small regions of high
star-forming activity have been found within 50~pc from the galactic
center that may contribute up to 1\% of the galactic star-formation
rate~\cite{Figer:2003tu}, although this finding does not seem to
contradict the overall picture of a reduced SN rate in the inner
galaxy.

\begin{figure}[b]
\centering
\includegraphics[width=0.55\textwidth]{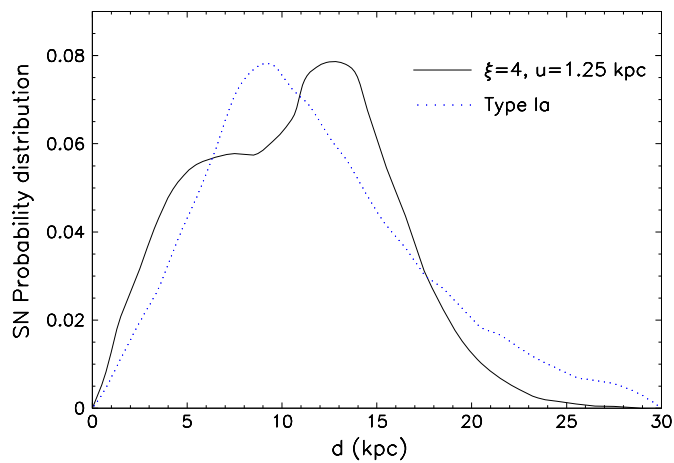}
\caption{\label{fig:distribution} Distance distribution of
core-collapse (solid) and thermonuclear SNe (dotted) according to
the assumed galactic surface distributions of Eqs.~(\ref{eq:param})
and (\ref{eq:distrTIa}), respectively~\cite{Mirizzi:2006xx}.}
\end{figure}

As a representative example we follow Ref.~\cite{Mirizzi:2006xx} and
consider a common parametrization for the galactic surface density
of core-collapse (cc) events,
\begin{equation}\label{eq:param}
{\sigma}_{\rm cc}(r)\propto r^\xi\exp(-r/u)\ ,
\end{equation}
where $r$ is the galactocentric radius.  For the birth location of
neutron stars, a fiducial distribution of this form was suggested
with the parameters $\xi=4$ and $u=1.25~{\rm kpc}$
~\cite{Yusifov:2004fr}. They are consistent with several SN-related
observables, even though large uncertainties remain. Thermonuclear
SNe, that are believed to originate from old stars in binary
systems, more closely follow the matter distribution. It can be
parameterized as~\cite{Ferriere:2001rg}
\begin{equation}\label{eq:distrTIa}
 {\sigma}_{\rm Ia}(r) \propto
 \exp\left(-\frac{r}{4.5~{\rm kpc}}\right)\,.
\end{equation}
We show the SN distance distributions corresponding to these models
in Fig.~\ref{fig:distribution}. The tails at large distances are
unphysical due to a complete lack of data.

The average distance for the assumed distribution is $\langle d_{\rm
cc}\rangle=10.7~{\rm kpc}$ with a rms dispersion of 4.9~kpc. This
agrees with the fiducial distance of 10~kpc that is frequently
assumed in the literature. On the other hand, the dispersion is very
large so that the number of neutrinos detected even from a
``typical'' galactic SN can vary by more than an order of magnitude.

\section{Basic lessons from a SN neutrino observation}
\label{sec:lessons}

\subsection{Early warning, distance and direction}

Turning to the many uses of a SN neutrino observation, we first note
that a it occurs several hours before the optical explosion,
allowing one to issue an alert. The Supernova Early Warning System
(SNEWS) provides this service to the neutrino and astronomy
communities~\cite{Antonioli:2004zb}.

Most galactic SNe are optically obscured. While it is implausible
that the SN will remain invisible in the entire electromagnetic
spectrum, it is interesting if it can be located by its neutrinos
alone~\cite{Beacom:1998fj, Tomas:2003xn}. The best existing pointing
capability is provided by $\nu+e\to\nu+e$ scattering in Super-K
where an accuracy of about $8^\circ$ (95\% CL half-cone opening
angle) can be achieved. If neutron tagging becomes possible by
adding gadolinium~\cite{Beacom:2003nk}, the accuracy increases to
about~$3^\circ$. For a megatonne-class detector with 30 times the
Super-K fiducial volume, these numbers improve to $1.4^\circ$ (no
neutron tagging) and $0.6^\circ$ (90\% tagging efficiency).

The distance of SN~1987A, besides is obvious association with the
Large Magellanic Cloud, could be directly determined with light
echoes from its inner ring~\cite{Panagia:1991, Gould:1997cc}. If the
next galactic SN is obscured, nothing of the sort may be possible
and one may actually have to rely on the neutrinos to estimate its
distance. However, SNe are no good neutrino standard candles. The
total emitted energy depends on the poorly known nuclear equation of
state as well as the total mass of the progenitor star. The signal
registered by the standard $\bar\nu_e+p\to n+e^+$ reaction is also
subject to details of the flavor-dependent neutrino emission and on
flavor oscillations. Altogether, one could probably estimate the
distance within a factor of two or so.

The prompt $\nu_e$ burst, on the other hand, comes close to being a
standard candle~\cite{Thompson:2002mw, Takahashi:2003rn,
Kachelriess:2004ds}. Here the problem is that the world lacks a big
$\nu_e$ detector because in water-Cherenkov and scintillator
detectors the main channel is inverse beta decay. In a large liquid
Argon TPC the charged-current absorption
$\nu_e+{}^{40}\hbox{Ar}\to{}^{40}\hbox{K}+e^-$ would provide an
exquisite $\nu_e$ signal~\cite{Gil-Botella:2003sz}. In a megatonne
water-Cherenkov detector with neutron tagging, the signal from $\nu+
e$ scattering could be isolated and a distance determination within
5--10\% may become possible, in particular if the neutrino mass
hierarchy and the 13-mixing angle were
known~\cite{Kachelriess:2004ds}.

\subsection{Neutrino spectrum}

The SN~1987A neutrino observations provided a unique confirmation of
the overall picture of core-collapse and neutron-star formation. The
signal lasted for about ten seconds, a time scale predicted by the
diffusive neutrino energy transport in a nuclear-density hot compact
star. The energies in the ten~MeV range, representative of the
temperature at the ``neutrino sphere,'' roughly agrees with
expectations. (The physics of core-collapse phenomena was presented
by H.-T.~Janka at this conference; for a recent review
see~\cite{Janka:2006fh}.)

In detail, however, the $\bar\nu_e$ energies implied by
Kamiokande-II~\cite{Hirata:1987hu, Hirata:1988ad} and
IMB~\cite{Bionta:1987qt, Bratton:1988ww} do not agree well with each
other or with expectations. In particular, the Kamiokande-II
energies are significantly lower than expected~\cite{Janka:1989,
Jegerlehner:1996kx, Mirizzi:2005tg, Costantini:2006xd}. To interpret
the SN~1987A data in any useful way one must make a prior assumption
about the spectral shape~\cite{Mirizzi:2005tg}. The tension in the
data and with theoretical models may well be a fluke of small-number
statistics, but a serious comparison of the neutrino spectrum with
theory for sure requires better data. Even a low-statistics signal
of a few tens of events from a SN in Andromeda in a megatonne
detector would provide valuable information. Without better data one
has to rely on theoretical models, for example, to interpret future
measurements of the cosmic Diffuse Supernova Neutrino Background
(DSNB) from all past SNe (see C.~Lunardini's presentation at this
conference~\cite{Lunardini:2006em}).

A large detector might reveal new subdominant spectral components. A
few 100--200~MeV events contemporaneous with the ordinary burst
could reveal that energy leaks out directly from the inner core in
some novel form of radiation. For example, right-handed neutrinos
produced in the SN core could decay into active
ones~\cite{Dodelson:1992tv} or neutrinos with Dirac magnetic moments
could escape from the SN interior and spin-precess into active ones
on the way to us~\cite{Barbieri:1988nh, Notzold:1988kz}.

\subsection{Signal duration}

The signal duration of the SN~1987A burst agrees well with
expectations. This observation is the basis for perhaps the most
useful particle-physics lesson from SN~1987A: apparently there was no
other energy-loss channel but the ordinary
neutrinos~\cite{Ellis:1987pk, Raffelt:1987yt, Turner:1987by,
Mayle:1987as}. This ``energy-loss argument'' has been applied to a
large number of cases, notably axions, Majorons, right-handed
neutrinos, and Kaluza-Klein gravitons, often providing the most
restrictive limits on the underlying particle-physics model.
Extensive reviews are Refs.~\cite{Schramm:1987ra, Raffelt:1990yz,
Raffelt:1999tx} and some more recent applications are discussed in
Refs.~\cite{Hanhart:2000er, Hanhart:2001fx, Dreiner:2003wh,
Fayet:2006sa}.

Far-reaching conclusions about fundamental physics are here based on
a sparse sample of data. Even a relatively low-statistics
observation would be enough to remove any lingering doubt if these
energy-loss limits are actually correct. Beyond a general
confirmation, a high-statistics observation would not improve such
limits very much because their uncertainties are typically dominated
by physics in the SN core. This includes uncertainties about the
temperature, density and composition of the medium as well as
uncertainties of how to calculate interaction and emission rates in
a nuclear medium.

\subsection{High-statistics light curve}

If one were to observe a high-statistics neutrino light curve,
crucial details of the core-collapse paradigm cold be tested. In
particular, one could probably separate the early accretion phase
from the later Kelvin-Helmholtz cooling phase after the explosion
has been launched. If the standard delayed-explosion scenario is
indeed correct, one could probably see the different phases in the
neutrino light curve and confirm or refute this
scenario~\cite{Totani:1997vj}. Besides Super-K, the IceCube detector
would be well suited to this task even though it does not provide
spectral information, but a high-statistics ``bolometric'' neutrino
light-curve that reflects the time-structure of the burst with high
significance.

A detailed cooling profile would allow one to test the theory behind
neutrino transport in a hot nuclear medium. Moreover, one may be
able to detect short-term time variations that are caused by the
large-scale convection pattern during the accretion phase. A sudden
termination would reveal late black-hole formation. Of course, there
could be completely unexpected features.

Even a high-statistics signal has only limited time-of-flight
sensitivity to neutrino masses. Even the most ambitious forecasts do
not seriously go below 1~eV~\cite{Totani:1998nf, Beacom:1998ya,
Nardi:2004zg}, not good enough in the light of cosmological
limits~\cite{Lesgourgues:2006nd, Hannestad:2006zg} and the expected
sensitivity of the KATRIN tritium decay
experiment~\cite{Drexlin:2005zt}. A few tens of events from a SN in
Andromeda would also provide a sensitivity of about 1~eV. One man's
trash is another man's treasure: we now expect the time-of-flight
dispersion caused by neutrino masses to be so small that fast time
variations at the source will faithfully show up at the detector.

\section{Neutrino flavor oscillations}
\label{sec:flavor-oscillations}

\subsection{Ordinary MSW oscillations}

Since SN~1987A, many of the ``simple'' questions about neutrinos
have been answered, but more challenges lie ahead. The observation
of a galactic SN burst may help us to address some of them. The
neutrinos pass through the mantle and envelope of the progenitor
star and encounter a vast range of matter densities, implying two
MSW resonances. One of them corresponds to the ``atmospheric mass
difference'' (H-resonance), the other, at lower density, to the
``solar mass difference'' (L-resonance). Of particular interest is
the MSW effect at the H-resonance driven by the unknown 13-mixing
angle. This resonance occurs in the neutrino sector for the normal
mass hierarchy, and among anti-neutrinos for the inverted hierarchy.
It is adiabatic for $\sin^2\Theta_{13}\agt 10^{-3}$ and
non-adiabatic for $\sin^2\Theta_{13}\alt 10^{-5}$. Therefore, the
neutrino burst is, in principle, sensitive to the mass hierarchy and
the 13-mixing angle~\cite{Dighe:1999bi, Dighe:2004xy}.

One important simplification is that the neutrino energies are far
below the $\mu$ and $\tau$ mass thresholds. Therefore, the species
$\nu_\mu$, $\bar\nu_\mu$, $\nu_\tau$, and $\bar\nu_\tau$ have only
neutral-current interactions. Their fluxes and spectra emerging from
the SN and their detection cross sections are the same. They are
collectively denoted as $\nu_x$ or equivalently $\bar\nu_x$. On the
other hand, $\nu_e$ and $\bar\nu_e$ have charged-current
interactions, notably with protons, neutrons and nuclei with
different abundances so that we finally need to distinguish between
the three species $\nu_e$, $\bar\nu_e$ and~$\nu_x$. Oscillation
effects can be summarized in terms of the energy-dependent $\nu_e$
survival probability $p(E)$ as
\begin{equation}
F_{\nu_e}(E) = p(E) F_{\nu_e}^0(E) + [1-p(E)] F_{\nu_x}^0(E)\,,
\label{pbar-def}
\end{equation}
where the superscript zero denotes the primary fluxes. An analogous
expression pertains to $\bar\nu_e$ with the survival probability
$\bar p(E)$. Table~\ref{tab:survival} summarizes the survival
probabilities for different mixing scenarios where $\Theta_\odot$
refers to the ``solar'' mixing angle~\cite{Dighe:1999bi,
Dighe:2004xy}.

\begin{table}[b]
\caption{\label{tab:survival}Survival probabilities for neutrinos,
  $p$, and antineutrinos, $\bar{p}$, in various mixing scenarios.
  The channels where one expects Earth effects, shock-wave propagation
  effects, and where the full $\nu_e$ burst is present or absent are
  indicated.}
\centering
\begin{tabular}{llllllll}
  \br
  Scenario&Hierarchy& $\sin^2\Theta_{13}$ & $p$ & $\bar{p}$&
  Earth effects&Shock wave&$\nu_e$ burst\\
  \mr
  A & Normal &${\agt}\,10^{-3}$  & 0  & $\cos^2\Theta_\odot$&
  $\bar\nu_e$& $\nu_e$&absent \\
  B & Inverted &  $\agt\, 10^{-3}$ &  $\sin^2\Theta_\odot$ &  0&
  $\nu_e$& $\bar\nu_e$&present\\
  C & Any & $\alt\, 10^{-5}$  & $\sin^2\Theta_\odot$ &
  $\cos^2\Theta_\odot$ &
  $\nu_e$ and $\bar\nu_e$&---&present\\
  \br
\end{tabular}
\end{table}

The most pronounced and most robust flavor-dependent structure of a
SN neutrino signal is the prompt $\nu_e$ burst. Unfortunately, the
main detection channel in all existing and near-future detectors is
$\bar\nu_e+p\to n+e^+$. In Super-K, the prompt $\nu_e$ burst would
generate of order 10 events from $\nu e$ scattering so that the
burst perhaps could be just barely detected. Of course, in a
megatonne water-Cherenkov detector with neutron tagging, the $\nu_e$
burst would be an extremely useful tool both for studying flavor
oscillations and determining the SN distance~\cite{Tomas:2003xn}.
Likewise, a large liquid Argon TPC would be a powerful and useful
$\nu_e$ detector~\cite{Gil-Botella:2003sz}.

For the time being, inverse beta decay will provide the dominant
signal. Oscillation effects are more subtle in this channel because
the primary spectra and fluxes of $\bar\nu_e$ and $\bar\nu_x$ are
probably more similar than had been thought until
recently~\cite{Keil:2002in, Raffelt:2003en}. Moreover, the relative
spectral energies and fluxes change during the accretion and cooling
phases. At present, reliable predictions for the time-dependent
quantities $\langle E_{\bar\nu_e}\rangle/\langle
E_{\bar\nu_x}\rangle$ and $F_{\bar\nu_e}/F_{\bar\nu_x}$ are not
available and in fact may differ for different SNe because the
progenitor mass may play some~role.

Therefore, one must focus on model-independent signatures. One is
the matter regeneration effect if the neutrinos are observed through
the Earth. Flavor oscillations would manifest themselves by
characteristic energy-dependent signal
modulations~\cite{Lunardini:2001pb, Lunardini:2003eh, Dighe:2003be,
Dighe:2003jg, Dighe:2003vm, Lindner:2002wm}, an effect that would be
especially apparent in a large scintillator detector because of its
superior energy resolution. One could also compare the signals of
different detectors if one of them sees the SN shadowed and the
other not~\cite{Dighe:2003be}. We have provided an online tool that
allows one, for chosen detector locations, to calculate the
probability for the next galactic SN to be shadowed in none, one, or
both detectors~\cite{Mirizzi:2006xx}. Both for SN and geo-neutrino
detection, several big scintillator detectors in different locations
would be more useful than one large detector such as the proposed
LENA~\cite{MarrodanUndagoitia:2006qs} in a single location.

Another characteristic signature of flavor oscillations could be a
pronounced dip or double-dip feature in the late neutrino signal
caused by shock-wave propagation. When the shock wave passes the
H-resonance region, the MSW adiabaticity is temporarily broken.
Moreover, for some time several H-resonances obtain because the
density profile is not monotonic. If one were to observe such
features, they could serve as a diagnostic both for neutrino
oscillation parameters and the astrophysics of shock-wave
propagation~\cite{Schirato:2002tg, Tomas:2004gr, Fogli:2004ff,
Choubey:2006aq}.

The SN matter profile need not be smooth. Behind the shock-wave,
convection and turbulence can cause significant stochastic density
variations that tend to wash out the neutrino oscillation
signatures~\cite{Fogli:2006xy, Friedland:2006ta}. The quantitative
relevance of this effect remains to be understood.

\subsection{Collective neutrino oscillations}

The trapped neutrinos in a SN core as well as the neutrinos
streaming off its surface are so dense that they provide a large
matter effect for each other. The nonlinear nature of this
neutrino-neutrino effect renders its consequences very different
from the ordinary matter effect in that it results in collective
oscillation phenomena~\cite{Pantaleone:1992eq, Samuel:1993uw,
Kostelecky:1993yt, Kostelecky:1993dm, Kostelecky:1994ys,
Kostelecky:1995dt, Kostelecky:1995xc, Samuel:1996ri,
Kostelecky:1996bs, Pantaleone:1998xi, Pastor:2001iu, Sawyer:2005jk,
Sigl:2007yz} that can be of practical interest in the early
universe~\cite{Lunardini:2000fy, Dolgov:2002ab, Wong:2002fa,
Abazajian:2002qx} or in core-collapse SNe~\cite{Pantaleone:1994ns,
Qian:1994wh, Sigl:1994hc, Pastor:2002we, Balantekin:2004ug,
Fuller:2005ae, Duan:2005cp, Duan:2006an, Duan:2006jv,
Hannestad:2006nj}. The crucial importance of ``bipolar
oscillations'' for SN neutrinos was first recognized in
Refs.~\cite{Duan:2005cp, Duan:2006an, Duan:2006jv} and some of their
salient features explained in Ref.~\cite{Hannestad:2006nj}.

What are the conditions for neutrino-neutrino matter effects to be
relevant? Considering for simplicity a two-flavor situation, vacuum
oscillations are driven by the frequency $\omega=\Delta m^2/2E$. The
ordinary matter effect is important when $\lambda\agt\omega$ where
$\lambda=\sqrt2 G_{\rm F} n_e$. Neutrino-neutrino effects are
important when $\mu\agt\omega$ where $\mu=\sqrt2 G_{\rm F} n_{\nu}$.
It is crucial to note that ordinary matter effects do not override
neutrino-neutrino effects. As stressed in Ref.~\cite{Duan:2005cp},
it is a misconception that neutrino-neutrino effects would be
negligible when $\lambda\gg\mu$.

The low-energy weak-interaction Hamiltonian is of current-current
form so that the interaction energy between two particles of momenta
${\bf p}$ and ${\bf q}$ is proportional to $(1-{\bf v}_{\bf
p}\cdot{\bf v}_{\bf q})$ where ${\bf v}_{\bf p}={\bf p}/E_{\bf p}$
is the velocity. In isotropic media the ${\bf v}_{\bf p}\cdot{\bf
v}_{\bf q}$ term averages to zero. On the other hand,
collinear-moving relativistic particles produce no weak potential
for each other. For neutrinos streaming off a SN core, the $(1-{\bf
v}_{\bf p}\cdot{\bf v}_{\bf q})$ term implies that the neutrino flux
declines not only with the geometric $r^{-2}$ factor, but the
average interaction energy $\mu$ has another $r^{-2}$ factor that
accounts for the increasing collinearity of the neutrino
trajectories with distance from the source~\cite{Qian:1994wh}.
Considering the atmospheric mass difference of
1.9--$3.0\times10^{-3}~{\rm eV}^2$ and using a typical energy of
15~MeV, we may use $\omega=0.3~{\rm km}^{-1}$ as a typical value,
where we here express frequencies and energies in km$^{-1}$ that is
a useful unit in the SN context. Moreover, if we use $10^{51}~{\rm
erg}~{\rm s}^{-1}$ as a typical neutrino luminosity, and if we use
10~km as the neutrino-sphere radius, we may use
$\mu=0.3\times10^5~{\rm km}^{-1}$ at the neutrino sphere so that
indeed $\mu\gg\omega$. With the $r^{-4}$ scaling of the effective
$\mu$, collective neutrino oscillations will be important out to a
radius of about 200~km.

There are two extreme cases of collective oscillation effects that
have been discussed in the literature. {\it Synchronized
oscillations\/} occur when the neutrino-neutrino interaction
``glues'' the neutrino flavor polarization vectors together enough
so that they evolve the same. In other words, even though the vacuum
oscillation frequency $\Delta m^2/2E$ is different for different
modes, they all oscillate with the same ``synchronized frequency''
that is an average of the vacuum or in-medium frequencies
(``self-maintained coherence''). Of course, if the vacuum or
in-medium mixing angle is small, this synchronization effect has no
macroscopic significance.

The generic case of {\it bipolar oscillations} occurs in a neutrino
gas with equal densities of, say, $\nu_e$ and $\bar\nu_e$. In an
inverted-mass situation with a small mixing angle, the ensemble will
undergo oscillations of the sort
$\nu_e\bar\nu_e\to\nu_\mu\bar\nu_\mu\to\nu_e\bar\nu_e\to\ldots$,
approximately with the ``bipolar frequency''
$\kappa=\sqrt{2\omega\mu}$ that is much faster than the vacuum
oscillation frequency. The period of this phenomenon depends
logarithmically on the mixing angle, explaining why this phenomenon
is not much affected by the presence of ordinary
matter~\cite{Duan:2005cp, Duan:2006an, Hannestad:2006nj}. For the
normal hierarchy, the ensemble performs small-amplitude harmonic
oscillations with the frequency $\kappa$ so that macroscopically
``nothing'' happens.

The next complication are ``multi-angle effects,'' probably first
stressed in Ref.~\cite{Sawyer:2005jk} and numerically explored in
Ref.~\cite{Duan:2006an}. In a non-isotropic neutrino gas, the
self-term is not the same for all modes because of the $(1-{\bf
v}_{\bf p}\cdot{\bf v}_{\bf q})$ factor. The result is an
instability that causes a neutrino gas with equal densities of $\nu$
and $\bar\nu$ to de-cohere kinematically in flavor space between
different directions of motion. Independently of the mass hierarchy
and with the smallest initial anisotropy, complete flavor
equipartition obtains. The time scale, again, is set by the bipolar
frequency $\kappa$. The overall time to achieve equilibrium depends
logarithmically on the mixing angle and the initial
anisotropy~\cite{Sigl:2007yz}.

Bipolar oscillations are a collective pair-conversion effect; there
is no enhanced flavor conversion. For equal densities of $\nu_e$ and
$\bar\nu_e$, the net electron lepton number vanishes. ``Pair
oscillations'' do not change of overall flavor lepton number. One
requirement is that there is a sufficient ``pair excess'' in some
flavor. This is not the case in the interior of a SN core where all
neutrinos are in thermal equilibrium, and only the $\nu_e$ have a
large chemical potential that increases the number density of
$\nu_e$ (relative to $\nu_\mu$ or $\nu_\tau$) while at the same time
suppressing the $\bar\nu_e$ density. Therefore, bipolar oscillations
do not seem to be relevant in the interior of a SN core.
Synchronized oscillations will occur, but with an extremely small
in-medium mixing angle.

On the other hand, there is an excess of both $\nu_e$ and $\bar\nu_e$ 
in the neutrinos streaming off a SN core, where generically
$F_{\nu_e}>F_{\bar\nu_e}$. If this asymmetry is too large, the
oscillations are still of the synchronized type, even though there
is a pair excess. Bipolar conversions will begin playing a role
beyond a radius where the effective $\mu$ is small enough that the
asymmetry no longer prevents them. The critical region is between a
few tens of km above the neutrino sphere and about 200~km. Without
the ``multi-angle effect,'' the outcome would be generic in that
complete pair-conversion $\nu_e\bar\nu_e\to \nu_x\bar\nu_x$ would
occur for the inverted mass hierarchy, and essentially nothing new
would happen for the normal hierarchy. Including multi-angle
effects, the outcome does not seem generic but rather depends on
details~\cite{Duan:2006an}.

As for observable flavor oscillation effects from the next galactic
SN, the deleponization burst likely remains unaffected because it is
characterized by an excess of $\nu_e$ and a suppression of
$\bar\nu_e$. During the accretion phase, some degree of
flavor-swapping may occur and since the relevant region is within the
stalled shock wave, one may speculate if some effect on the SN
dynamics itself obtain in the spirit of Ref.~\cite{Fuller:1992}.
After a successful explosion, nucleosynthesis in the neutrino-driven
wind above the neutron star may well be affected, a possibility that
was the main motivation for the exploratory study of
Ref.~\cite{Duan:2006an, Qian:1993dg}. Possible modifications of what
will be observed in the neutrino signal of the next galactic SN have
not yet been studied. Some interesting work remains to be done!

\section{Summary}
\label{sec:summary}

Twenty years after SN~1987A we are well prepared for the observation
of another neutrino burst from a collapsing star. The scientific
harvest would be immense. Without any doubt, neutrinos would be
excellent astrophysical messengers and allow us to follow stellar
collapse and many of its details ``in situ.'' From the
particle-physics perspective, many of the unique lessons from
SN~1987A could be corroborated. In principle, the neutrino burst
also holds information about the neutrino mass hierarchy that is
extremely difficult to determine in the laboratory. On the other
hand, collective neutrino oscillation effects that had not been
fully appreciated may change some of the previous paradigm. In
preparation for the next galactic SN burst, both theorists and
experimentalists have more work to do than just wait!

\ack {This work was partly supported by the Deutsche
Forschungsgemeinschaft under Grants No.~SFB-375 and TR-27 and by the
European Union under the ILIAS project, contract
No.~RII3-CT-2004-506222.}

\section*{References}

\end{document}